\documentclass{article}[11pt]
\usepackage{amsfonts}
\usepackage{booktabs}
\usepackage{graphicx}
\usepackage{CJK,amsmath,indentfirst}
\usepackage{amssymb}
\usepackage{amsmath}
\usepackage{amsthm}
\usepackage{epsfig}
\usepackage{times}
\usepackage{subfigure}
\usepackage[numbers,sort&compress]{natbib}
\begin{document}

\title{\bf Interaction solutions for (1+1)-dimensional higher-order Broer Kaup system }

\author{\footnotesize  Xiangpeng Xin  \thanks{Corresponding author ~~~~ Email:xinxiangpeng2012@gmail.com} \\
\footnotesize  School of Mathematical Sciences, Liaocheng University, Liaocheng 252059, People's Republic of China}
\date{}
\maketitle
\parindent=0pt
\textbf{Abstract:}  The (1+1)-dimensional higher-order Broer-Kaup (HBK) system is studied by consistent tanh expansion (CTE) method in this paper.It is proved that the HBK system is CTE solvable. Some exact interaction solutions among different nonlinear excitations such as solitons, rational waves, periodic waves and corresponding images are explicitly given. 

\vskip.4in
\renewcommand{\thesection}{\arabic{section}}
\parindent=20pt

\section{Introduction}
Finding explicit solutions of nonlinear evolution equations (NLEEs) is one of the most important problems in mathematical physics. Some effective methods have been introduced, such as Painlev\'{e} analysis\cite{Weiss1,Conte1}, classical and non-classical Lie group\cite{Olver1,Bluman1}, nonlocal symmetry\cite{Lou1,Galas1,Hu1}, variable separation approach\cite{Lou2,Lou3}, and various function expansion methods\cite{Fan1,Fan2} etc. Using these methods, researchers obtained several types of exact solutions, including interaction solutions. However, it is very difficult to find interaction solutions among different types of nonlinear excitations except for soliton-soliton interactions.

It is known that Painlev\'{e} analysis is an important method to investigate the integrable property of a given NLEE, and the truncated Painlev\'{e} expansion method is a straight way to provide auto-B\"{a}cklund transformation and analytic solution, furthermore, it can also be used to obtain nonlocal symmetries. Recently, by developing the truncated Painlev\'{e} expansion, Lou\cite{Lou4,Lou5,Lou6} defined a new integrability in the sense of possessing a consistent tanh expansion. This method is greatly valid for constructing various interaction solutions between different types of excitations. For example, solitons, cnoidal waves, Painlev\'{e} waves, Airy waves, Bessel waves etc. It has been revealed that many more integrable systems are CTE solvable and posses quite similar interaction solutions which can be described by the same determining equation with different constant constraints.

In this work, we will consider the following (1+1)-dimensional higher-order Broer-Kaup (HBK) system\cite{Lou7},
\begin{equation}\label{eq1}
\begin{array}{l}
 u_t = - 4(u_{xx}  + u^3  + 6uv - 3uu_x )_x , \\
 v_t = - 4(v_{xx}  + 3uv_x  + 3u^2 v + 3v^2 )_x , \\
 \end{array}
\end{equation}
where $u = u(x,t),v = v(x,t)$, The HBK system (\ref{eq1}) and the following (1+1)-dimensional Broer-Kaup system
\begin{equation}\label{eq2}
\begin{array}{l}
 u_t  = u_{xx}  - 2uu_x  - 2v_x , \\
 v_t  =  - v_{xx}  - 2(uv)_x , \\
 \end{array}
\end{equation}
can be obtained from the Kadomtsev-Petviashvili equation. For the system (\ref{eq2}), some exact interaction solutions among different nonlinear excitations such as solitons, rational waves, periodic waves, error function waves and any Burgers waves are explicitly given\cite{Lou8}. System (\ref{eq1}) can be seen as an extension of the known Broer-Kaup system which is often used to model the bi-directional propagation of long waves in shallow water. The Painlev\'{e} integrability and auto-B\"{a}cklund transformation were presented in\cite{Li1}. N-fold Darboux transformation and multi-soliton solutions were given in\cite{Huang1}. However, to our knowledge, the interaction solutions between the solitons and other different types of nonlinear waves for system (\ref{eq1}) have not been
studied in literatures.

This paper is arranged as follows: In Sec.2, the consistent tanh expansion(CTE) method is introduced. In Sec.3, Using the CTE method, some interaction solutions between different types of excitations and corresponding images are given. Finally, some conclusions and discussions are given in Sec.4.

\section{Consistent tanh expansion method}

For a given derivative nonlinear system,
\begin{equation}\label{eq3}
F({\rm{x}},t,u) = 0,{\rm{x}} = (x_1 ,x_2 ,...),
\end{equation}
assumes that the system (\ref{eq3}) has following possible truncated expansion solution
\begin{equation}\label{eq4}
u = \sum\limits_{i = 0}^n {u_i \tanh ^i (w)},
\end{equation}
where $w$ is an undetermined function of space time $x$ and $t$, $n$ should be determined from the leading order analysis of Eq.(\ref{eq3}).

Substituting Eq.(\ref{eq4}) into Eq.(\ref{eq3}), one can obtain all the expansion coefficient functions by vanishing the coefficients of powers tanh(w). $w$ will be obtained by vanishing all the coefficients of powers tanh(w) or, not over-determined.

 \textbf{Definition 1:} If $u_i $ $(i = 0, . . . , n) $ and $w$ are obtained for the system (\ref{eq3}), then the expansion (\ref{eq4}) is called a CTE and the nonlinear system (\ref{eq3}) is CTE solvable.

\section{Interaction solutions for the HBK system}

By using the leading order analysis for the HBK system (\ref{eq1}), we can take the following truncated tanh expansions

\begin{equation}\label{eq5}
\begin{array}{l}
 u = u_0  + u_1 \tanh (w), \\
 v = v_0  + v_1 \tanh (w) + v_2 \tanh ^2 (w), \\
 \end{array}
\end{equation}

Substituting Eq.(\ref{eq5}) into Eq.(\ref{eq1}), vanishing coefficients of all the powers of tanh(w), there are eleven over-determined equations for only six undetermined functions $u_0,u_1,v_0,v_1,v_2$ and $w$. solving the over-determined system and obtained the following solution,
\begin{equation}\label{eq6}
\begin{array}{l}
 u_0  =  - \frac{{w_{xx} }}{{w_x }},u_1  = 2w_x , \\
 v_0  = \frac{{32w_x^4  - 16w_x^3 w_x  - w_x w_t  + 12w_{xx}^2 }}{{24w_x^2 }}, \\
 v_1  = 2w_{xx} ,v_2  =  - 2w_x^2 , \\
 \end{array}
\end{equation}
and the function $w$ satisfies
\begin{equation}\label{eq7}
16w_t w_x^4  - 8w_x^2 w_{xxt}  + 16w_x w_{xx} w_{xt}  + w_x w_t^2  - 4w_t w_{xx}^2  = 0,
\end{equation}
so, we have the following CTE theorem for the HBK system.

\textbf{Theorem 1.} If $w$ is a solution to the equation(\ref{eq7}), then
\begin{equation}\label{eq8}
\begin{array}{l}
 u =  - \frac{{w_{xx} }}{{w_x }} + 2w_x \tanh (w), \\
 v = \frac{{32w_x^4  - 16w_x^3 w_x  - w_x w_t  + 12w_{xx}^2 }}{{24w_x^2 }} + 2w_{xx} \tanh (w) - 2w_x^2 \tanh ^2 (w), \\
 \end{array}
\end{equation}
constitute a CTE solution of the HBK system (\ref{eq1}).

According to theorem 1, one can obtain various interaction solutions among different types of nonlinear excitations by solving the Eq. (\ref{eq7}).

\textbf{Case 1.} Resonant soliton solution.

It is not difficult to verify that Eq.(16) possesses the following wave solution

\begin{equation}\label{eq9}
w = \frac{1}{4}\ln (\frac{{c_2 (\sqrt {c_2 } \tan (\Delta ) + c_3 )^2 }}{{4(c_3 c_5 \sqrt {c_2 }  - 4c_4 c_1^{3/2}  + c_5 \sqrt {c_2 } \tan (\Delta ))^2 }}),
\end{equation}
where $\Delta  = \frac{{\sqrt {c_2 } (c_1 x + c_2 t + c_3 )}}{{4c_1^{3/2} }}$, and $c_1,c_3,c_4,c_5$ are arbitrary constants $c_2\geq0$.

Substituting Eq. (\ref{eq9}) into the CTE (\ref{eq8}), one can obtain resonant soliton solutions of the HBK system which displays soliton fission and fusions. Here omitting specific expression of $u$ and $v$ because they are too long. In order to study the properties of the solutions, we choose some parameter values and draw the corresponding images using maple software which can be seen in appendix, and the parameters used in the figure
are selected as $c_1=0.5,c_2=0.5,c_3=0.5,c_4=0.8,c_5=0.1$.

\textbf{Remark 1:} Figure 1 displays this kind of resonant soliton solutions for the fields $u$ and $v$. One can see from the images that the amplitude of interaction wave are changing over time, which can be seen clearly from density figures with contour plot shown. The solutions displays soliton fission and fusions which can be easily applicable to the analysis of physically interesting processes for example the generation process of ghost waves.

\textbf{Case 2}: Soliton interactions with periodic waves.

Interactions between a soliton and a cnoidal periodic wave can be clarified by a common formula with different dispersion relations. One can take the $w$ function as some kinds of Jacobi elliptic functions, Here, just take $w$ in the following special simple form,
\begin{equation}\label{eq10}
w = k_0 x + l_0 t + \lambda {\rm{EllipticF}}(sn(k_1 x + l_1 t,m),m),
\end{equation}
$k_0,k_1,l_0,l_1,\lambda$ and $m$ are determined later, and EllipticF(z,k) is incomplete elliptic integral of the first kind,
\begin{center}
${\rm{EllipticF}}(z,k) = \int_0^z {\frac{1}{{\sqrt {1 - \alpha ^2 } \sqrt {1 - \alpha ^2 k^2 } }}d\alpha },$
\end{center}
with the help of the Maple, substituting Eq.(\ref{eq10}) into Eq.(\ref{eq9}), solving the over-determined equations and obtained the following solution
\begin{equation}\label{eq11}
l_0  =  - 16\lambda ^3 k_1^3  - 48\lambda ^2 k_0 k_1^2  - 48\lambda k_1 k_0^2  - 16k_0^3  - \lambda l_1 ,
\end{equation}
$k_0,k_1,l_1,m,\lambda$ are arbitrary constants.

Substituting Eqs.(\ref{eq10}),(\ref{eq11}) into the CTE (\ref{eq8}), soliton interactions with periodic waves of the HBK system can be obtained. Because the result is very prolix, here omitting it. Corresponding images can be seen in appendix and the parameters used in the figure
are selected as $k_1=30,\l_1=0.1,m=0.95,\lambda=-0.01,k_0=0.9$.

\textbf{Remark 2:} Figure 2 displays this kind of soliton-cnoidal wave solutions for the fields $u$ and $v$, including the interaction of the kink soliton+cnoidal periodic wave for $v$ and the soliton+cnoidal periodic wave for $u$ respectively. This kinds of solutions describe a soliton propagates on a cnoidal wave background for the HBK equation. If setting the module $m=1$, the soliton-cnoidal wave interaction solution reduces back to the two-soliton solution.

\section{Discussion and Summary}

In summary, by developing the truncated Painlev\'{e} analysis, we using the CTE method to solve a well-known dispersive water wave system, the HBK system. It is found that the HBK system is not only integrable under some nonstandard meaning but also CTE solvable. Some interaction solutions among solitons and other types of nonlinear waves which may be explicitly expressed by the Jacobi elliptic functions and the corresponding elliptic integral are constructed by means of the CTE method.

To search for interaction solutions of integrable DEs are considerable interest and value, the consistent tanh expansion provides a simple and effective way.One can find more interaction solutions for other CTE solvable systems using this method. Moreover, nonlocal symmetries and corresponding group invariant solutions can be constructed by the truncated Painlev\'{e} analysis and CTE method. Above topics will be discussed in the future series research works.

\section*{Acknowledgments}

This work is supported by National Natural Science Foundation of China under Grant (Nos.11505090, 11405103, 11447220),Research Award Foundation for Outstanding Young Scientists of Shandong Province (No.BS2015SF009).

\section*{Appendix}

%
%
%

\newpage

\small{
}
\end{document}